 \documentstyle[prl,aps]{revtex}  

\begin{document}
\draft
\preprint{\today}


 \twocolumn[\hsize\textwidth\columnwidth\hsize  
 \csname @twocolumnfalse\endcsname              

\title{Anomalous rotational-alignment in N=Z nuclei and 
       residual neutron-proton interaction} 
\author{Yang Sun$^{1,2}$ and Javid A. Sheikh$^{3}$  
        }
\address{
$^1$Department of Physics and Astronomy, University of Tennessee,
Knoxville, Tennessee 37996\\
$^2$Department of Physics, Xuzhou Normal University,
Xuzhou, Jiangsu 221009, P.R. China\\
$^3$Physik-Department, Technische
Universit\"at M\"unchen, D-85747 Garching, Germany 
}

\date{\today}
\maketitle

\begin{abstract}
Recent experiments have demonstrated that the rotational-alignment
for the $N=Z$ nuclei in the mass-80 region is considerably delayed as compared
to the neighboring $N \ne Z$ nuclei. 
We investigate whether this observation can be understood by a known component
of nuclear residual interactions.  
It is shown that the quadrupole-pairing interaction, which explains many of
the delays known in rare-earth nuclei, does not produce the 
substantial delay observed for these $N=Z$ nuclei.
However, the residual neutron-proton interaction which is conjectured
to be relevant for $N=Z$ nuclei is shown to be quite important
in explaining the new experimental data. 
\end{abstract}


 ]  

\narrowtext
\bigskip

The major advancements in the $\gamma$-ray detecting systems has made it possible 
for the first time to measure the high-spin states of the $N=Z$ nuclei,
$^{72}$Kr, $^{76}$Sr and $^{80}$Zr \cite{Fi01,Kel01}, to a spin region where the
rotational-alignment of nucleons is normally expected. A comparison of
the new data with the neighboring $N\ne Z$ nuclei shows that the alignment
is consistently and considerably delayed for these $N=Z$ nuclei. This delay cannot be
explained \cite{Ang97} using the cranked shell model 
approach which has been a powerful tool for studying the high-spin phenomena. 
  
The proton-rich mass-80 nuclei exhibit many phenomena that are quite unique
to this mass region. Unlike the mid-rare earths and actinides that have a very
stable deformation, the structure changes are quite pronounced among the
neighboring nuclei in the mass-80 region.
Indeed, recent experiment \cite{Kr74} found much reduced transition quadrupole moments
above the alignment process in $^{74}$Kr,
indicating a dramatic shape change along the
yrast sequence. 

For the $N\approx Z$ nuclei,
there has been an open question whether the neutron-proton (n-p)
correlations play a role in the structure analysis.
The existence of n-p pairing in nuclear systems, in particular
in nuclei with equal number of neutrons and protons, is one 
of the topical issues. It has been demonstrated using the cranking 
approaches in a single-$j$ shell \cite{Frau99,Zhang98,Sheikh99,Wyss00}
that the rotational-alignment 
properties will be modified by the residual n-p
interaction. In particular, it has been shown that the rotational-alignment 
of nucleons will be delayed for $N=Z$ nuclei as compared to the neighboring
$N \ne Z$ nuclei. However, the models used in these studies are very 
schematic and the results of these calculations cannot be compared with the
experimental data. It is also not very apparent from these
calculations which component of the n-p residual interaction is responsible
for the delay in the rotational-alignment.
 
Recently, we have performed a systematic investigation for the
yrast properties of the proton-rich Kr, Sr and Zr nuclei \cite{Pa01}.  
The analysis was carried out using the projected shell model (PSM) 
approach \cite{review}. 
We calculated the $N=Z, Z+2$ and $Z+4$ isotopes and studied their moments
of inertia, transition quadrupole moments and g-factors. Quantitative  
comparisons were made with available data, and predictions were given
for quantities that were not measured.  
The variations along the isotopic and isotonic chains were attributed 
to the mixing of various configurations of the projected deformed
Nilsson states as a function of shell filling.

A comparison of our PSM calculations in ref. \cite{Pa01} for
the three Kr-isotopes $^{72,74,76}$Kr
is shown in Fig. 1, now including the new data of $^{72}$Kr \cite{Fi01}. 
It is evident from the figure that, although, a reasonable  
agreement can be seen for $^{74}$Kr \cite{Kr74} and $^{76}$Kr \cite{Kr76} 
isotopes, a clear 
disagreement in the plot of moment of inertia occurs for 
the $N=Z$ nucleus $^{72}$Kr. 
The calculated moment of inertia for the yrast states of $^{72}$Kr shows a 
sharp backbend which is in contradiction with the smooth
curve obtained experimentally. The experimental moment of inertia 
indicates a slight upbend around $\hbar \omega =0.8$ MeV. The rotational-alignment 
is therefore significantly delayed as compared to the neighboring $^{74,76}$Kr
isotopes. 
It is known that the rotational-alignment may be delayed in cases with larger
deformation, but this argument cannot be applied here since both
the theoretical calculations and the experimental low-spin transition energies 
suggest that the deformation 
in $^{72}$Kr does not differ very much from those of the neighboring $^{74,76}$Kr. 

The purpose of the present paper is to explore how the behavior of
moment of inertia for $^{72}$Kr can be understood in the PSM 
framework. In particular, we would like to see whether the observation 
can be explained by a known type
of nuclear residual interactions.
As mentioned earlier, the single-$j$ shell models indicate that the
n-p residual interaction may be responsible for the delayed rotational-alignment
observed in $N=Z$ nuclei. We would like to investigate the behavior of the
moment of inertia as a function of the strength of the n-p interaction
employed in the PSM approach.

The description of the PSM calculations for Kr-, Sr- and Zr-isotopes can be found 
in ref.\cite{Pa01}. We would just like to mention that in the PSM calculations
for these nuclei,  
three major shells ($N=2,3,4$) for both neutron and proton are used
and the shell model space includes the 0-, 2- and 4-quasiparticle (qp) states:
\begin{equation}
\left|\phi \right>_\kappa  = \left\{\left|0 \right>, \ 
\alpha^\dagger_{n_i} \alpha^\dagger_{n_j} \left|0 \right>,\
\alpha^\dagger_{p_i} \alpha^\dagger_{p_j} \left|0 \right>,\
\alpha^\dagger_{n_i} \alpha^\dagger_{n_j} \alpha^\dagger_{p_i}
\alpha^\dagger_{p_j} \left|0 \right> \right\} ,
\label{baset}
\end{equation}
where $\alpha^\dagger$ is the creation operator for a qp and the
index $n$ ($p$) denotes neutron (proton) Nilsson quantum numbers which run over
the low-lying orbitals below the cut-off energy. 
The deformed single particles are generated by the Nilsson calculation
\cite{tod}. 
It is important to note 
that for the $N=Z$ nuclei under consideration, unperturbed 2-qp 
states of $\alpha^\dagger_{n_i} \alpha^\dagger_{n_j} \left|0 \right>$
and $\alpha^\dagger_{p_i} \alpha^\dagger_{p_j} \left|0 \right>$ 
with the same configuration can occur pairwise as nearly
degenerate states.

The Hamiltonian employed in the PSM calculation 
contains the separable forces and 
can be expressed as $\hat
H = \hat H_\nu + \hat H_\pi + \hat H_{\nu\pi}$, where
$H_\tau$ $(\tau=\nu,\pi)$ is the like-particle pairing plus quadrupole
Hamiltonian, with the inclusion of quadrupole-pairing,
\begin{equation}
\hat {H}_{\tau}~=~\hat H^0_{\tau} -{\chi_{\tau\tau}\over
2} \sum_\mu \hat Q^{\dagger\mu}_{\tau} \hat
Q^{\mu}_{\tau} - G_{\mbox{\scriptsize M}}^{\tau}\hat
P^\dagger_{\tau} \hat P_{\tau} - G_{\mbox {\scriptsize
Q}}^{\tau} \sum_\mu \hat P^{\dagger\mu}_{\tau} \hat
P^{\mu}_{\tau},
\label{E}
\end{equation}
and $\hat H_{\nu\pi}$ is the n--p
quadrupole--quadrupole residual interaction
\begin{equation}
\hat H_{\nu\pi} ~=~ - \chi_{\nu\pi} \sum_\mu \hat
Q^{\dagger \mu}_\nu \hat Q^\mu_\pi.
\label{QQ}
\end{equation}
In Eq.\ (\ref{E}), the four terms are,
respectively, the spherical single-particle energy, the
quadrupole-quadrupole interaction, the monopole-pairing,
and the quadrupole-pairing interaction. The interaction
strengths
$\chi_{\tau \tau}$ ($\tau = \nu$ or $\pi$) are related
self-consistently to the deformation $\varepsilon_2$ by
\begin{equation}
\chi_{\tau \tau} ~=~ {{{2\over 3}
\varepsilon_2 (\hbar\omega_\tau )^2} \over
{\hbar\omega_\nu \langle \hat Q_0 \rangle_\nu +
\hbar\omega_\pi \langle \hat Q_0 \rangle_\pi}}.
\label{chi}
\end{equation}
Following ref.\ \cite{review}, the strength
$\chi_{\nu\pi}$ is assumed to be 
\begin{equation}
\chi_{\nu\pi} ~=~
(\chi_{\nu\nu} \chi_{\pi\pi})^{1/2}. 
\label{chinp}
\end{equation}
Similar
parameterizations were used in many of the earlier works 
\cite{PQmodel}.

The interaction strengths in Eq.\ (\ref{E}) are consistent 
with the values used in the previous PSM calculations
for the same mass region \cite{Pa01,Do98}.
The monopole pairing strength $G_M$ is taken to be
$G_M=\left[18.0-14.5(N-Z)/A\right]/A$ for neutrons and $G_M=14.5/A$ for
protons. 
The quadrupole pairing
strength $G_Q$ is assumed to be $G_Q=\gamma G_M$, the
proportionality constant $\gamma$  
being fixed to 0.16.

The eigenvalue equation of the PSM for a given spin $I$ takes the
form \cite{review}
\begin{equation}
\sum_{\kappa'}\left\{H^I_{\kappa\kappa'}-E^IN^I_{\kappa\kappa'}\right\}
F^I_{\kappa'}=0.
\label{psmeq}
\end{equation}
The expectation value of the Hamiltonian with respect to a ``rotational-
band $\kappa$'', $H^I_{\kappa\kappa}/N^I_{\kappa\kappa}$ defines  
an unperturbed band energy and when plotted as functions of spin $I$, 
it is called a band diagram \cite{review}. 
Valuable information can often be extracted from the unperturbed bands
before the configuration mixing. For example, crossing of two
unperturbed bands near the yrast line is usually of particular interest. 
The results after band mixing depend on how the bands cross 
(crossing with large or small crossing angles) and how they
interact with each other (types of residual interactions) 
at the crossing region. 

Band-crossings can be visualized in a band diagram, for example,  
those between the
ground-band (g-band or 0-qp band) with 2-qp bands usually consisting
of the (decoupled) high-$j$ particles. 
Delayed rotational-alignment may be related to unusual types of
interactions. 
The n-p pairing is thought to be relevant here because one is  
dealing with the $N=Z$ systems.   
However, the quadrupole pairing is also a type of 
residual pairing interaction used in 
nuclear structure calculations \cite{QP}, and  
it has been known that this force can delay a rotational- alignment 
process \cite{WF78}.

In ref. \cite{SWF94}, it was demonstrated that the 
quadrupole pairing force in the Hamiltonian of Eq.~(\ref{E}) is crucial
to shift the rotational-alignment in the rare earth region. 
For the case $^{175}$Ta, 
by increasing the strength constant  
$\gamma$ from 0.16 to 0.24, a pronounced delay of rotational-alignment was 
obtained, which successfully explained the data. 
Ref. \cite{SF96} further showed the systematic results for a large number of
Lu, Ta, Re and
Ir isotopes, with the overall agreement with data being very satisfactory.
Generally, the larger the delay of crossings,
the larger the quadrupole pairing force. Thus, these results \cite{SWF94,SF96}
suggested that
from the study of rotational-alignment, one could extract information about
the quadrupole pairing
interaction in the rotating systems.

In the present study, we would like to investigate whether similar calculations
with adjusting the quadrupole pairing force can reproduce the observed
alignment delay in $N=Z$ nuclei. 
In Fig. 2, we show the results for $^{72}$Kr with various quadrupole pairing
strengths $\gamma$.
In the calculations,  
$\gamma$ is allowed to vary from 0.16 to 0.28, and 
$\gamma=0.16$ was the original value used in calculations in ref. \cite{Pa01}.
It can be seen that 
we have obtained a delay in crossing frequency, 
but the amount of delay is too small to explain the data.
It should also be emphasized here that a sharp backbend in the plot remains
for all different values of $\gamma$, in a total disagreement with the new data. 
Thus, we conclude that 
the quadrupole pairing effect is not the primary source for the delayed
alignment in $^{72}$Kr. 
 
Next, we would like to study the effect of the neutron-proton interaction.
As already mentioned, several theoretical calculations suggest
that the residual n-p interaction may be important in the discussion
of the high-spin phenomena in $N=Z$ nuclei. 
Specifically for $N=Z$ nuclei, the pairwise occurrence of neutron
and proton 2-qp bands in the band-crossing region suggests 
the neutron-proton type interaction can be very important for these nuclei.
Together with the g-band, 
it constitutes a band-crossing picture involving three bands.
This is in contrast with the common picture of two-band crossing
for most cases in $N\ne Z$ nuclei.
The sharp backbend obtained in 
the PSM approach using the standard parameters
may suggest that the neutron-proton interaction used is too weak.  
Here, we would like to investigate
in a purely phenomenological manner the influence of the
n-p interaction, through the n-p $QQ$ term 
(see Eq. (\ref{QQ})) in the PSM, on the behavior
of the moment of inertia. 
It should be noted that the strengths 
of the proton-proton and neutron-neutron $QQ$, $\chi_{\pi\pi}$ and
$\chi_{\nu\nu}$ in Eq. (\ref{E}), are still determined by the
self-consistency condition, Eq.(\ref{chi}), and we allow only changes 
in the n-p strength $\chi_{\nu\pi}$. 
The standard strength of the n-p $QQ$ given by
Eq. (\ref{chinp}) is based on the assumption of the iso-scalar coupling.
The implication of our treatment here is that 
this assumption may not be valid in general, and may be
modified by the residual n-p interaction. 

The calculations for three $N=Z$ nuclei $^{72}$Kr, $^{76}$Sr and $^{80}$Zr
are presented in Fig. 3. For each of these nuclei, 
we increase the strength $\chi_{\nu\pi}$ by
multiplying a factor $\alpha=1.1$, 1.2 and 1.3. 
A rather pronounced effect can be seen for $^{72}$Kr: 
the increase of the n-p interaction has a combined effect  
of delaying crossing frequency and smoothing the backbending curve. 
Eventually, the basic feature in the data can be described 
when a strong  n-p interaction is used.  
On the other hand, effect of decreasing $\chi_{\nu\pi}$ is also shown in Fig. 3.
Calculations with $\alpha=0.8$ give the rotational- alignments
at a lower frequency.

The two-fold effect of a stronger n-p interaction can be clearly seen 
in the band diagram. When using the factor $\alpha=1.3$, the rotational
behavior of both neutron
and proton 2-qp bands are modified so that they cross the g-band 
at a higher spin and with a smaller crossing angle.
In addition, the stronger n-p force acts between the bands resulting in a
smoother behavior for the yrast states. 
We would like to mention that a stronger $\chi_{\nu\pi}$ which
reproduces the new data for $^{72}$Kr will offset
the good agreement previously obtained for the $N\ne Z$ nuclei $^{74,76}$Kr  
with the standard interaction strength. 

For the other two $N=Z$ nuclei $^{76}$Sr and $^{80}$Zr in Fig. 3,
it can been seen that the effect  
of the n-p interaction for the band-crossing region  
seems to have an isotonic dependence. 
It is observed that increasing the strength $\chi_{\nu\pi}$ by the 
same amount for $^{76}$Sr and $^{80}$Zr does not lead to the same 
pronounced effect as in $^{72}$Kr. In fact, varying the factor $\alpha$ from 1.0 to 1.3,
we obtain similar smooth curves as our original one \cite{Pa01} for 
$^{76}$Sr, which has already reproduced data well. 

It is remarkable that for $^{72}$Kr these rather different rotational-alignment
features are obtained as a result of different emphasis on the
neutron-neutron, proton-proton, and neutron-proton interactions.  
Therefore empirically increasing the strength of the n-p $QQ$ interaction
appears to mimic the experimental trends and simulates the expected 
enhanced correlations in the np-channel for these $N=Z$ nuclei.

In principle, the residual n-p interaction which is
supposed to be important for the $N=Z$ systems should be of pairing
type. It has been known from several mean-field studies (see, for example, \cite{Goo79})
that the n-p pairing
is non-zero for $N=Z$ nuclei and vanishes for $N \ne Z$ nuclei. However, the
n-p pairing contains pairs of higher angular-momenta \cite{Wyss00}
apart from $J=1$ pairs. These higher angular-momentum pairs will have a 
significant particle-hole contribution and may modify, for example, 
the $QQ$ interaction used in the PSM Hamiltonian. Therefore, increasing
the strength of the $QQ$ in the n-p interaction term, as has been shown
in the present work, has a physical origin.
It is interesting to see here that the n-p pairing term
tends to renormalize the $QQ$ interaction for $N=Z$ nuclei only, 
since for $N \ne Z$
nuclei any changes will destroy the good agreement obtained 
with the standard interaction strength.  

In conclusion, the new experimental data on the $N=Z$ nuclei 
in the mass-80 region have  
clearly indicated that the rotational-alignment is
substantially delayed for these nuclei. This delay cannot be explained
using the standard interaction in the projected shell model
approach, which has worked well in the high-spin description for
a broad range of nuclei. The quadrupole-pairing interaction, which explains many of
the delay in the rotational-alignment for $N\ne Z$ nuclei, does not produce the
substantial delay observed in the present $N=Z$ examples. 
It has been shown that the n-p residual interaction may be
responsible for this delay. Our calculations have clearly demonstrated 
that increasing the
strength of the n-p interaction term in the two-body Hamiltonian 
significantly delays the rotational-alignment.
However, to make a quantitative comparison with the experimental
data, the renormalization of the interaction for the $N=Z$ nuclei
has to be obtained by employing explicitly the n-p pairing term
as a residual part in the Hamiltonian.
This work is now under progress and the results will be published
in the future.

The authors wish to express their thanks to Dr. C.J. Lister 
for stimulating discussions and a careful reading of this manuscript with 
valuable suggestions.
 
\baselineskip = 14pt
\bibliographystyle{unsrt}

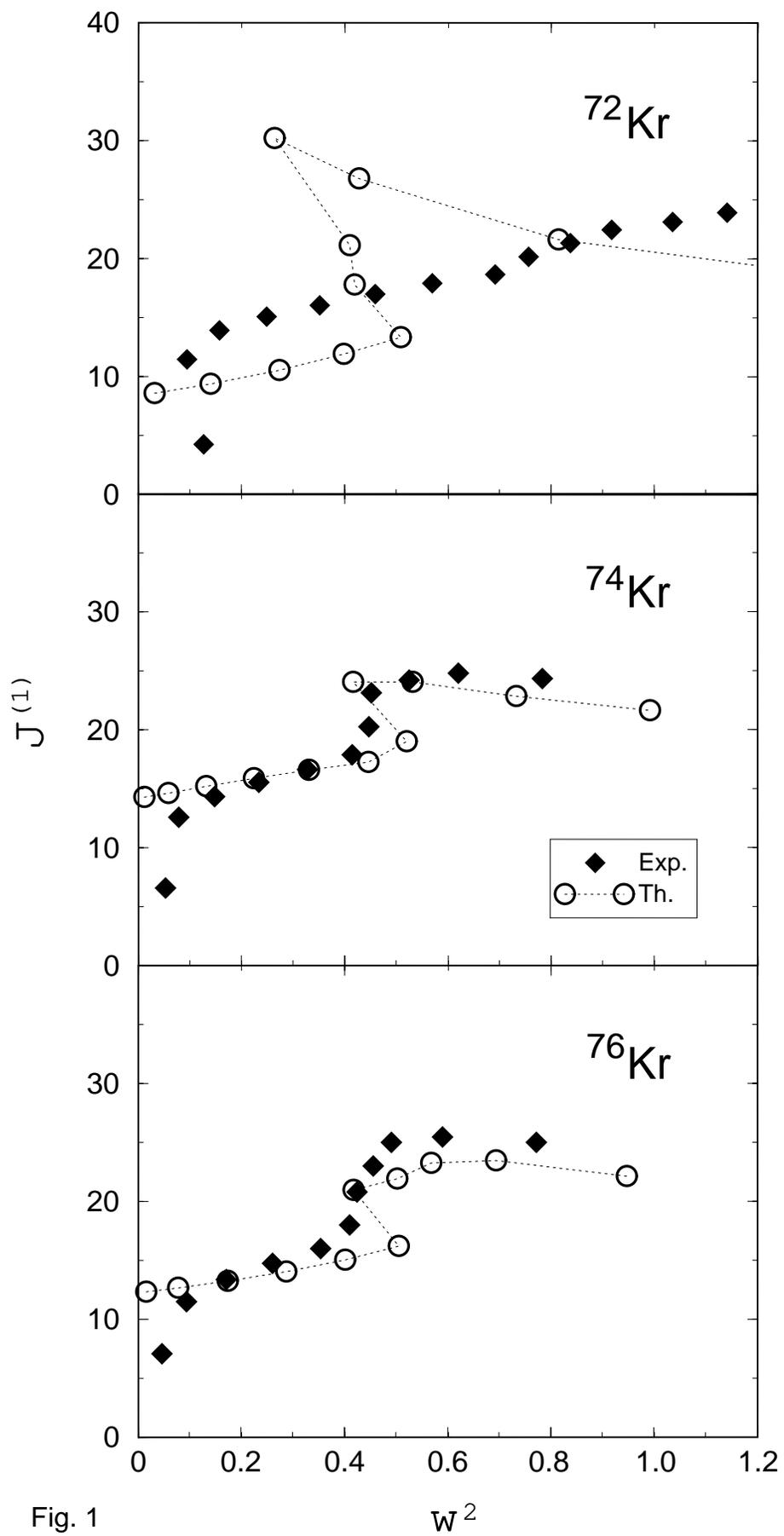
\begin{figure}
\caption{
Moments of inertia $\Im^{(1)}={{2I-1}\over {E(I)-E(I-2)}} (\hbar^2/MeV)$
as function of $\omega^2$ with $\omega={{E(I)-E(I-2)}\over 2} (MeV/\hbar)$. 
Theoretical results with the standard interaction in the PSM are
compared with the experimental data. 
The experimental data are taken from ref. 
\protect\cite{Fi01} for $^{72}$Kr, ref. \protect\cite{Kr74} for $^{74}$Kr, and
ref. \protect\cite{Kr76} for $^{76}$Kr. 
}
\label{figure.1}
\end{figure}

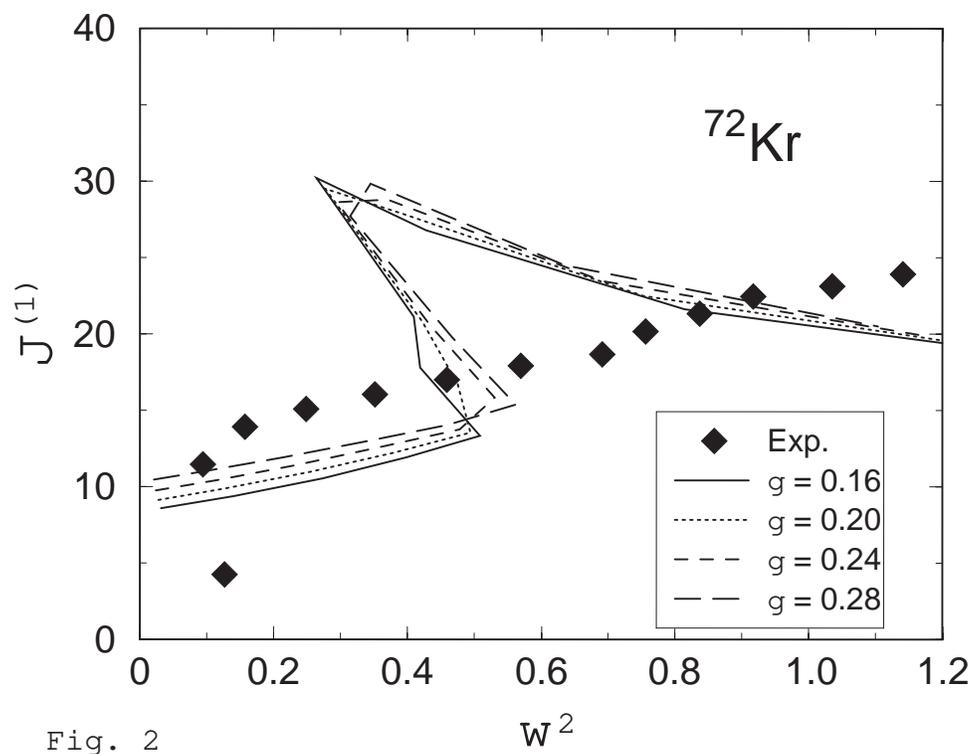
\begin{figure}
\caption{ 
Moments of inertia $\Im^{(1)}={{2I-1}\over {E(I)-E(I-2)}} (\hbar^2/MeV)$
as function of $\omega^2$ with $\omega={{E(I)-E(I-2)}\over 2} (MeV/\hbar)$. 
Theoretical results with various quandrupole pairing strengths are
compared with the $^{72}$Kr data. 
The experimental data are taken from ref. 
\protect\cite{Fi01}. 
}
\label{figure.2}
\end{figure}

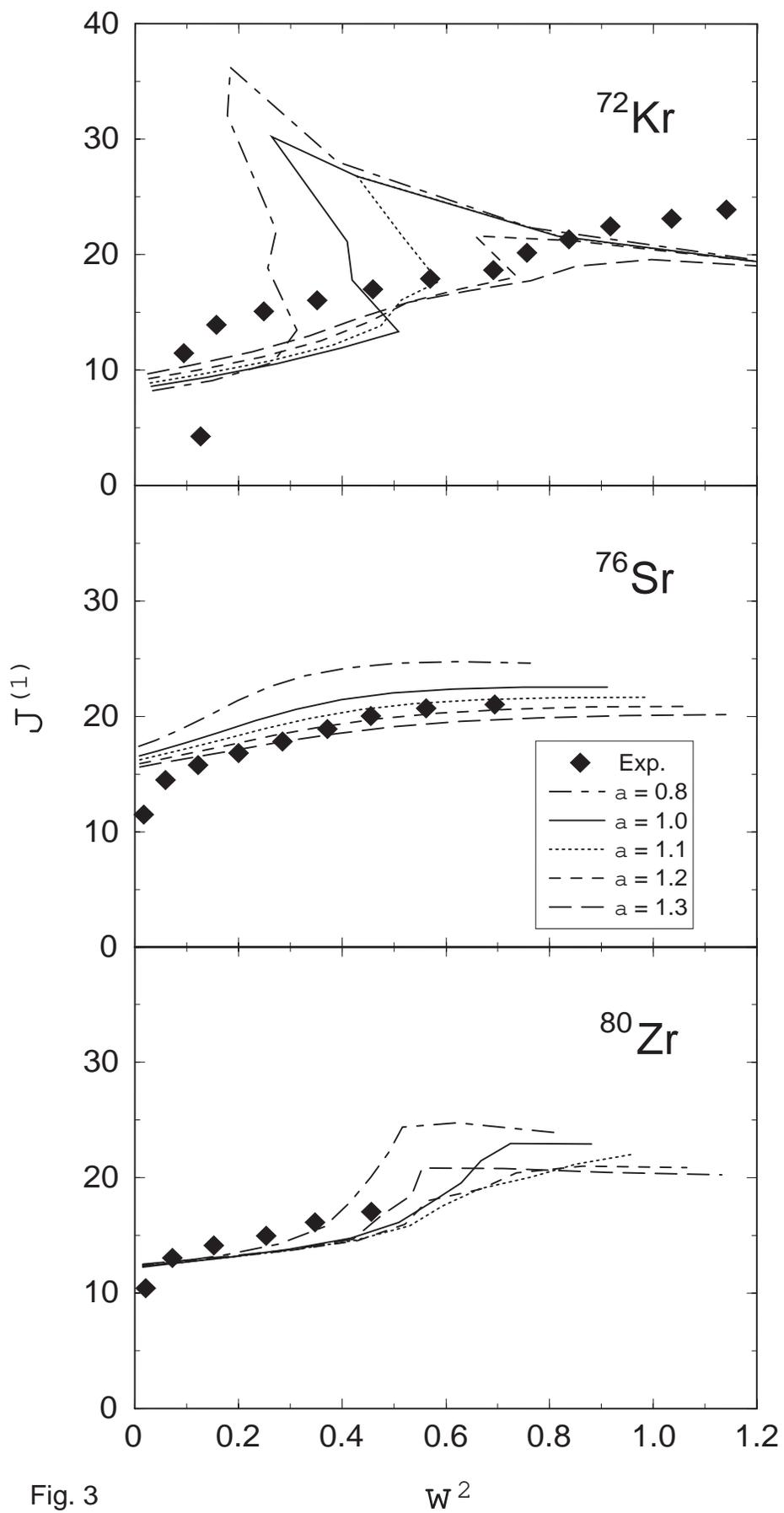
\begin{figure}
\caption{ 
Moments of inertia $\Im^{(1)}={{2I-1}\over {E(I)-E(I-2)}} (\hbar^2/MeV)$
as function of $\omega^2$ with $\omega={{E(I)-E(I-2)}\over 2} (MeV/\hbar)$. 
Theoretical results with various n-p interaction strengths are
presented together with the experimental data for three $N=Z$ nuclei,
$^{72}$Kr, $^{76}$Sr and $^{80}$Zr.
The experimental data are taken from ref.
\protect\cite{Fi01}.
}
\label{figure.3}
\end{figure}

\end{document}